\documentclass[prd,eqsecnum,preprint,showpacs]{revtex4}
\usepackage{bm}
\usepackage{stmaryrd}
\usepackage{amsmath}
\usepackage{amssymb}
\usepackage{graphicx}
\usepackage{textcomp}
\usepackage{calrsfs}
\usepackage{yfonts}

\newcommand{\be}{\begin{equation}}
\newcommand{\ee}{\end{equation}}
\newcommand{\ben}{\begin{equation*}}
\newcommand{\een}{\end{equation*}}
\newcommand{\bea}{\begin{eqnarray}}
\newcommand{\eea}{\end{eqnarray}}

\DeclareMathOperator{\Tr}{Tr}
\DeclareMathOperator{\tr}{tr}

\begin{document}

\title{Casimir energy, dispersion, and the Lifshitz formula}

\date{\today}

\author{Kimball A. Milton}\email{milton@nhn.ou.edu}
\author{Jef Wagner}\email{wagner@nhn.ou.edu}
\author{Prachi Parashar}\email{prachi@nhn.ou.edu}
\affiliation{Oklahoma Center for High Energy Physics and Homer L. Dodge
 Department of
Physics and Astronomy, University of Oklahoma, Norman, OK 73019-2061, USA}
\author{Iver Brevik}\email{iver.h.brevik@ntnu.no}
\affiliation{Department of Energy and Process Engineering, Norwegian University
of Science and Technology, N-7491 Trondheim, Norway}

\begin{abstract}
Despite suggestions to the contrary, we show in this paper that the usual
dispersive form of the electromagnetic energy must be used to derive the
Lifshitz force between parallel dielectric media.  This conclusion follows
from the general form of the quantum vacuum energy, which is the basis of
the multiple-scattering formalism.  As an illustration, we explicitly 
derive the Lifshitz formula for the interaction between parallel dielectric
semispaces, including dispersion, starting from the expression for the
total energy of the system. The issues of constancy of the energy
between parallel plates and of the observability of electrostrictive forces
are briefly addressed.

\end{abstract}

\maketitle
\section{Introduction}
Recent years have yielded considerable progress in understanding quantum
vacuum or Casimir energies, both theoretically and experimentally.
For a very recent review see Ref.~\cite{Bordagbook}.  However, there are
controversial aspects, both having to do with the concept of zero-point energy
applied to a single system, or to the universe as a whole \cite{Jaffe}, and
with including thermal corrections, and their observability in experiment
\cite{Bordagbook}.
The latter question refers to how the electric properties of materials 
depend on (imaginary) frequencies, that is, upon dispersion.

In this paper, we address the latter issue.  In a recent paper \cite{brevik08}
we had proposed, following a suggestion of Lifshitz \cite{landaulifshitz},
that the usual dispersive term in the electromagnetic energy for a given
frequency \cite{schwingerbook},
\be
U=\frac12\int (d\mathbf{r})
\int_{-\infty}^\infty \frac{d\omega}{2\pi}
\left[\frac{d(\omega\varepsilon)}{d\omega}E^2(\mathbf{r})
+H^2(\mathbf{r})\right]\label{dispen}
\ee
(we ignore the magnetic susceptibility, that is, we set $\mu=1$), should not
be included.  However, the usual derivations of the Lifshitz interaction
between dielectric slabs are not based on the total energy.  For example, in
Ref.~\cite{miltonbook} the Lifshitz formula is derived from the pressure,
or equivalently the spatial components of the stress tensor, and also from
the variational principle enunciated in Ref.~\cite{sdm}.  It is also easy to
obtain this same result using the recently repopularized multiple-scattering
approach to Casimir energies \cite{renne,reynaud}.  Equivalently, the 
multiple-reflection expansion yields the Lifshitz formula immediately
\cite{miltonrev}.

In this note, we derive the Lifshitz energy directly from Eq.~(\ref{dispen}).
We first see, in Sec.~\ref{sec1},
how dispersion is incorporated in a general formulation.  This
demonstrates that the dispersive form of the energy is required.  Then, 
after giving the form of the Green's dyadic in Sec.~\ref{sec2}, in
Sec.~\ref{sec3} we will explicitly derive the Lifshitz formula from 
Eq.~(\ref{dispen}), and will see manifestly that the dispersive term provides
the Jacobian of the required transformation of coordinates necessary to
obtain the necessary $\log\det$ form.  In the Conclusions, we also bring
up the related possibility of measuring electrostrictive effects in liquids.
The Appendix points out that the well-known constancy of the energy density
between parallel perfectly conducting plates does not hold for dielectric
plates (that is, if regions 1 and 2, defined in Sec.~\ref{sec2}, are
constituted of dielectric material),
or even if a dispersive medium exists between metallic plates.

\section{General formulation}
\label{sec1}
Let us start from Eq.~(\ref{dispen}), and consider the quantum vacuum energy
associated with electromagnetic field fluctuations:
\be
\mathfrak{E}=\frac12 \int (d\mathbf{r})\int_{-\infty}^\infty 
\frac{d\omega}{2\pi}
\left[\frac{d(\varepsilon\omega)}{d\omega}\langle E^2\rangle+\langle H^2\rangle
\right].\label{vev}
\ee
The expectation values appearing here are given by the electromagnetic
Green's dyadic,
\begin{subequations}
\bea
\langle\mathbf{E(r)E(r')}\rangle&=&\frac1i \bm{\Gamma}(\mathbf{r,r'}),\\
\langle\mathbf{H(r)H(r')}\rangle&=-&\frac1i\frac1{\omega^2}
\bm{\nabla}\times \bm{\Gamma}(\mathbf{r,r'})\times
\overleftarrow{\bm{\nabla}'},
\eea
\end{subequations}
and so inserting these into the energy expression (\ref{vev}), integrating
by parts, and using the differential equation satisfied by the Green's
dyadic,
\be
-\bm{\nabla}\times\bm{\nabla}\times\bm{\Gamma}+\omega^2\varepsilon\bm{\Gamma}=-
\omega^2\bm{1},\label{diffeq}
\ee
we obtain the expression for the energy
\be
\mathfrak{E}=-\frac{i}{2}\int(d\mathbf{r})\int\frac{d\omega}{2\pi}
\left[2\varepsilon\tr\bm{\Gamma}+
\omega\frac{d\varepsilon}{d\omega}\tr\bm{\Gamma}\right].\label{dispenergy}
\ee

We can obtain this same result starting from the standard trace-log formula:
\bea
\mathfrak{E}&=&\frac{i}2\int\frac{d\omega}{2\pi}\Tr\ln\bm{\Gamma}
=-\frac{i}2\int\frac{d\omega}{2\pi}\omega\frac{d}{d\omega}\Tr\ln\bm{\Gamma}
\nonumber\\
&=&-\frac{i}2\int\frac{d\omega}{2\pi}\omega\Tr\bm{\Gamma}^{-1}
\frac{d}{d\omega}\bm{\Gamma}=\frac{i}2\int\frac{d\omega}{2\pi}\omega
\Tr\bm{\Gamma}\frac{d}{d\omega}\bm{\Gamma}^{-1}\nonumber\\
&=&\frac{i}2\int\frac{d\omega}{2\pi}\omega\Tr\left(-\frac{2}
{\omega^3}\bm{\nabla}\times\bm{\nabla}\times-\frac{d\varepsilon}
{d\omega}\right)
\bm{\Gamma}=-\frac{i}2\int\frac{d\omega}{2\pi}\Tr\left(2\varepsilon\bm{\Gamma}
+\omega\frac{d\varepsilon}{d\omega}\bm{\Gamma}\right),\label{trlog}
\eea
where from Eq.~(\ref{diffeq}) 
\be
\bm{\Gamma}^{-1}=\frac1{\omega^2}\bm{\nabla}\times\bm{\nabla}\times-
\varepsilon,
\ee
and where $\Tr$ includes the trace over spatial coordinates.
The final form in Eq.~(\ref{trlog}) 
is exactly the result (\ref{dispenergy}) derived from the expectation
value of the classical electromagnetic energy (\ref{dispen}).

To conclusively demonstrate that the dispersive term must be included,
we derive the variational principle used to obtain the Lifshitz formula
in Ref.~\cite{sdm}.  This depends upon the variational statement
\be
\delta\bm{\Gamma}=-\bm{\Gamma}\delta\bm{\Gamma}^{-1}\bm{\Gamma}=\bm{\Gamma}
\delta\varepsilon\bm{\Gamma}.
\ee  Also, using the differential equation for the Green's dyadic we find
\be
\frac{d\bm{\Gamma}}{d\omega}=-\bm{\Gamma}\frac{d\bm{\Gamma}^{-1}}{d\omega}
\bm{\Gamma}=\frac2\omega\bm{\Gamma}\varepsilon\bm{\Gamma}+\bm{\Gamma}\frac{d
\varepsilon}{d\omega}\bm{\Gamma}+\frac2\omega\bm{\Gamma}.
\ee
Therefore, the $\varepsilon$-variation of Eq.~(\ref{dispenergy}) yields
\bea
\delta\mathfrak{E}&=&-\frac{i}2\int\frac{d\omega}{2\pi}\Tr\left(2\delta
\varepsilon
\bm{\Gamma}+2\varepsilon\bm{\Gamma}\delta\varepsilon\bm{\Gamma}+\omega\frac{d
\delta\varepsilon}{d\omega}\bm{\Gamma}
+\omega\frac{d\varepsilon}{d\omega}\bm{\Gamma}
\delta\varepsilon\bm{\Gamma}\right)\nonumber\\
&=&-\frac{i}2\int\frac{d\omega}{2\pi}\omega\frac{d}{d\omega}
\Tr\delta\varepsilon\bm{\Gamma},
\eea
which, upon integration by parts, yields the variational principle used in
Refs.~\cite{sdm,miltonbook}:
\be
\delta\mathfrak{E}=\frac{i}2\int\frac{d\omega}{2\pi}\Tr\delta
\varepsilon\bm{\Gamma}.
\ee
See also Ref.~\cite{brevikfest}.

\section{Green's dyadic for parallel slabs}
\label{sec2}
In this and the following section we supply an explicit derivation of the 
Casimir-Lifshitz interaction between parallel dielectric slabs (of infinite
thickness).  Specifically, consider a dielectric function in the following
form
\be
\varepsilon(\mathbf{r})=\left\{\begin{array}{cc}
\epsilon_1,&z<0,\\
\epsilon_3,&0<z<a,\\
\epsilon_2,&a<z.
\end{array}\right.\label{dielectricconst}
\ee
Then the Green's dyadic can be written as a transverse Fourier transform,
\be
\bm{\Gamma}(\mathbf{r,r'})=\int\frac{(d\mathbf{k_\perp})}{(2\pi)^2} 
e^{i\mathbf{k_\perp\cdot(r-r')_\perp}}\mathbf{g}(z,z';\mathbf{k_\perp},\omega),
\ee
where the reduced Green's dyadic has the form \cite{sdm,miltonbook},
\be
\mathbf{g}(z,z')=\left(\begin{array}{ccc}
\frac1\varepsilon\frac\partial{\partial z}\frac1{\varepsilon'}\frac\partial
{\partial z'}g^E&0&\frac{ik}{\varepsilon\varepsilon'}\frac\partial{\partial z}
g^E\\
0&\omega^2g^H&0\\
-\frac{ik}{\varepsilon\varepsilon'}\frac\partial{\partial z'}g^E&0&
\frac{k^2}{\varepsilon\varepsilon'}g^E\end{array}\right).
\ee
Here we have dropped $\delta$-function terms, we have denoted
 $\varepsilon=\varepsilon(z)$, $\varepsilon'=\varepsilon(z')$, and we
have chosen the coordinate system so that $\mathbf{k_\perp}=k\hat x$.  Here
the TE ($H$) and TM ($E$) 
(relative to the $z$ axis) Green's functions satisfy the differential equations
\begin{subequations}
\bea
\left(-\frac{\partial^2}{\partial z^2}+\kappa^2\right)g^H&=&\delta(z-z'),\\
\left(-\frac{\partial}{\partial z}\frac1\varepsilon\frac{\partial}{\partial z}
+\frac1\varepsilon\kappa^2\right)g^E&=&\delta(z-z').
\eea
\end{subequations}
We will solve these equations in each of the three regions given in
Eq.~(\ref{dielectricconst}), subject to boundary conditions between the
regions that $g^H$ and $\partial_z g^H$ are continuous, and that
$g^E$ and $(1/\varepsilon)\partial_z g^E$ are continuous.  These boundary
conditions reflect the underlying requirement that the transverse parts
of $\mathbf{E}$ and $\mathbf{H}$ are continuous, while the normal component
of $\mathbf{D}=\varepsilon\mathbf{E}$ is continuous (there are no surface
charges or currents).  It is a straightforward calculation to find the Green's
functions in each region.  We display the results for the only situation
we need in the following, when $z$ and $z'$ are both in the same regions.
Below the first interface, $z,z'<0$,
\begin{subequations}
\bea
g^H(z,z')&=&\frac1{2\kappa_1}\left[e^{-\kappa_1|z-z'|}+r_1 e^{\kappa_1(z+z')}
\right],\\
r_1&=&\frac{\kappa_1-\kappa_3}{\kappa_1+\kappa_3}+\frac{4\kappa_1\kappa_3}
{\kappa_3^2-\kappa_1^2}\frac1d,
\eea
\end{subequations}
where $\kappa^2_a=k^2+\zeta^2\varepsilon_a(i\zeta)$, $a=1,2,3$, and we have
made a Euclidean rotation, $\omega=i\zeta$.  Here we have introduced the 
abbreviation
\be
d=\frac{\kappa_3+\kappa_2}{\kappa_3-\kappa_2}\frac{\kappa_3+\kappa_1}
{\kappa_3-\kappa_1}e^{2\kappa_3a}-1.
\ee
Similarly, above the second interface, $z,z'>a$,\begin{subequations}
\bea
g^H(z,z')&=&\frac1{2\kappa_2}\left[e^{-\kappa_2|z-z'|}
+r_2 e^{-\kappa_2(z+z'-2a)}\right],\\
r_2&=&\frac{\kappa_2-\kappa_3}{\kappa_2+\kappa_3}+\frac{4\kappa_2\kappa_3}
{\kappa_3^2-\kappa_2^2}\frac1d.
\eea
\end{subequations}
In the intermediate region, $a>z,z'>0$,
\bea
g^H(z,z')&=&\frac1{2\kappa_3}\left[e^{-\kappa_3|z-z'|}
+\frac2d\cosh\kappa_3(z-z')\right.\nonumber\\
&&\qquad\mbox{}+\left.\frac{\kappa_3+\kappa_1}{\kappa_3-\kappa_1}
\frac1d e^{\kappa_3(z+z')}
+ \frac{\kappa_3+\kappa_2}{\kappa_3-\kappa_2}\frac1d e^{-\kappa_3(z+z'-2a)}
\right].\label{intgreen}
\eea

The transverse magnetic Green's function $g^E$ is obtained from the above by
replacing $\kappa_a\to\kappa_a/\epsilon_a$ except in the exponents.

The fact that $g(z,z)$ depends on $z$ implies, in general, that the 
mean-squared electric and magnetic fields also depend on position, as does
the energy density.  This seems to contradict the fact that for parallel
conducting plates the energy density is constant in each region \cite{brown,
brevik08}.  We shall show, in fact, in the
Appendix that the electromagnetic energy between
perfectly conducting plates is indeed constant, provided the intervening medium
is nondispersive.
\section{Lifshitz energy}
\label{sec3}
The Casimir-Lifshitz energy per unit area
 for the situation of parallel slabs described by
the dielectric function (\ref{dielectricconst}) is \cite{miltonbook,Bordagbook}
\be
\mathcal{E}=\frac1{4\pi^2}\int_0^\infty d\zeta\int_0^\infty dk\, k\left[
\ln\left(1-r_{\rm TE}r_{\rm TE}'e^{-2\kappa_3 a}\right)+
\ln\left(1-r_{\rm TM}r_{\rm TM}'e^{-2\kappa_3 a}\right)\right].\label{lifshitz}
\ee
where
\be 
r_{\rm TE}=\frac{\kappa_3-\kappa_1}{\kappa_3+\kappa_1},\quad
r'_{\rm TE}=\frac{\kappa_3-\kappa_2}{\kappa_3+\kappa_2},
\ee
and the TM reflection coefficients are obtained by replacing $\kappa_a\to
\kappa'_a=\kappa_a/\epsilon_a$.  In this section, we rederive this result from
Eq.~(\ref{dispenergy}).  

\subsection{TE contribution to the energy}
The TE part of the energy can be written as
\be
\mathcal{E}^{\rm TE}=\frac1{4\pi^2}\int_0^\infty d\zeta\int_0^\infty k\,dk
\int_{-\infty}^\infty dz\, \zeta\frac{d}{d\zeta}(-\kappa^2) g^H(z,z).
\ee
In each region, the dispersive term is necessary to change variables from
$\zeta$ to $\kappa_a$.  In the first region, omitting infinite terms which
contain no reference to the separation $a$ between the regions, we find
rather immediately (we assume, as usual, $\zeta^2\epsilon_a(\zeta)\to0$ as
$\zeta\to0$)
\be
\mathcal{E}^{\rm TE,1}=-\frac1{2\pi^2}\int_0^\infty k\,dk\int_k^\infty 
d\kappa_1 \frac{\zeta}2\frac\partial{\partial\kappa_1}\ln \Delta_{\rm TE},
\ee
where the partial derivative means that $\kappa_2$ and $\kappa_3$ are not
altered, and
\be
\Delta_{\rm TE}=1-r_{\rm TE}r_{\rm TE}'e^{-2\kappa_3 a}.
\ee
Similarly, in region 2,
\be
\mathcal{E}^{\rm TE,2}=-\frac1{2\pi^2}\int_0^\infty k\,dk\int_k^\infty
d\kappa_2 \frac{\zeta}2\frac\partial{\partial\kappa_2}\ln \Delta_{\rm TE}.
\ee
The intermediate region involves a slightly more involved calculation, but
the result has the same form (after we omit a constant term in the force):
\be
\mathcal{E}^{\rm TE,3}=-\frac1{2\pi^2}\int_0^\infty k\,dk\int_k^\infty
d\kappa_3 \frac{\zeta}2\frac\partial{\partial\kappa_3}\ln \Delta_{\rm TE},
\ee
where now the derivative acts also 
on the exponent in $\Delta_{\rm TE}$.  In this
way we obtain exactly the expected TE contribution:
\bea
\mathcal{E}^{\rm TE}&=&-\frac1{2\pi^2}\int_0^\infty k\,dk\int_k^\infty
d\kappa_3\frac\zeta2\left(\frac{\partial}{\partial\kappa_3}+
\frac{d\kappa_2}{d\kappa_3}\frac\partial{\partial\kappa_2}+
\frac{d\kappa_1}{d\kappa_3}\frac\partial{\partial\kappa_1}\right)\ln
\Delta_{\rm TE}\nonumber\\
&=&\frac1{4\pi^2}\int_0^\infty dk\,k\int_0^\infty d\zeta\ln\Delta_{\rm TE},
\eea
which is just the first term in Eq.~(\ref{lifshitz}).

\subsection{TM contribution to energy}
The TM contribution requires a somewhat more elaborate calculation.
The TM contribution to the trace of the Green's dyadic is
\be
\tr \mathbf{g}^{E}=\frac1\varepsilon\frac\partial{\partial z}
\frac1{\varepsilon'}\frac{\partial}{\partial z'}g^E+\frac{k^2}{\varepsilon
\varepsilon'}g^E.
\ee
This differential structure has different forms depending on whether it acts on
the pure exponential terms in $g^E$, or on the hyperbolic cosine in 
Eq.~(\ref{intgreen}), namely, in the first case,
\be
\frac1\varepsilon\frac\partial{\partial z}
\frac1{\varepsilon'}\frac{\partial}{\partial z'}+\frac{k^2}{\varepsilon
\varepsilon'}\to\frac1{\epsilon_a^2}(2
\kappa_a^2-\zeta^2\epsilon_a),
\ee
and in the second case,
\be
\frac1\varepsilon\frac\partial{\partial z}
\frac1{\varepsilon'}\frac{\partial}{\partial z'}+\frac{k^2}{\varepsilon
\varepsilon'}\to-\frac{\zeta^2}{\epsilon_3}.
\label{special}
\ee
Except for that last exceptional case, combining this trace term with the
dispersive term in the energy gives
\be
\left(\epsilon_a+\frac\zeta2\frac{d\epsilon_a}{d\zeta}\right)
\frac1{\epsilon_a^2}\left(2\kappa_a^2-\zeta^2\epsilon_a\right)=
2\kappa_a\left(\kappa_a'-\frac\zeta2\frac{d\kappa_a'}{d\zeta}\right).
\ee
Thus, in region 1, the contribution to the TM energy is
\be
\mathcal{E}^{\rm TM,1}=\frac1{4\pi^2}\int_0^\infty k\,dk\int_0^\infty
d\zeta\left(\kappa'_1-\frac\zeta 2\frac{d\kappa_1'}{d\zeta}\right)
\frac\partial{\partial\kappa_1'}\ln\Delta_{\rm TM},
\ee
where $\Delta_{\rm TM}$ differs from $\Delta_{\rm TE}$ by replacing $\kappa_a$
by $\kappa_a'=\kappa_a/\epsilon_a$ except in the exponents, that is, 
$r_{\rm TE}\to r_{\rm TM}$.
Similarly, in region 2,
\be
\mathcal{E}^{\rm TM,2}=\frac1{4\pi^2}\int_0^\infty k\,dk\int_0^\infty
d\zeta\left(\kappa'_2-\frac\zeta 2\frac{d\kappa_2'}{d\zeta}\right)
\frac\partial{\partial\kappa_2'}\ln\Delta_{\rm TM},
\ee
In region 3, however, we have to take into account the special case 
(\ref{special}).  It is most convenient then to regard $\kappa_3$ and
$\kappa_3'$ as independent, in which case we can write
\be
\mathcal{E}^{\rm TM,3}=\frac1{4\pi^2}\int_0^\infty k\,dk\int_0^\infty
d\zeta\left[\left(\kappa'_3-\frac\zeta 2\frac{d\kappa_3'}{d\zeta}\right)
\frac\partial{\partial\kappa_3'}-\frac\zeta2\frac{d\kappa_3}{d\zeta}
\frac\partial{\partial\kappa_3}\right]
\ln\Delta_{\rm TM}.
\ee
Thus, the total TM contribution is
\bea
\mathcal{E}^{\rm TM}&=&\frac1{4\pi^2}\int_0^\infty dk\,k\left\{
\int_0^\infty d\zeta\left[\kappa_1'\frac\partial{\partial\kappa_1'}
+\kappa_2'\frac\partial{\partial\kappa_2'}+
\kappa_3'\frac\partial{\partial\kappa_3'}\right]\ln\Delta_{\rm TM}
\right.\nonumber\\
&&\quad\mbox{}-\left.\int_k^\infty d\kappa_3\frac\zeta2
\left(\frac{\partial}{\partial\kappa_3}+
\frac{d\kappa'_3}{d\kappa_3}\frac\partial{\partial\kappa'_3}+
\frac{d\kappa'_2}{d\kappa_3}\frac\partial{\partial\kappa'_2}+
\frac{d\kappa'_1}{d\kappa_3}\frac\partial{\partial\kappa'_1}\right)\ln
\Delta_{\rm TM}\right\}.
\eea
The first term here is actually zero, because the differential operator
annihilates $\Delta_{\rm TM}$, since
\be
\kappa'_1\frac\partial{\partial\kappa'_1} r_{\rm TM}=-
\kappa'_3\frac\partial{\partial\kappa'_3} r_{\rm TM},
\ee
and so we obtain exactly the Lifshitz result
\be
\mathcal{E}^{\rm TM}=\frac1{4\pi^2}\int_0^\infty k\,dk\int_0^\infty d\zeta
\ln\Delta_{\rm TM}.
\ee

\section{Conclusions}
Ordinarily one calculates the Casimir-Lifshitz free energy directly from 
the pressure, or from an equivalent variational approach.  Therefore, it was
not obvious how the dispersive term present in the energy in order to have
the required balance between energy and momentum, as in the electromagnetic
energy-momentum tensor, plays a role.  Earlier we had suggested \cite{brevik08}
that such a term simply be omitted.  However, we now see that the dispersive term is
precisely what is needed to achieve agreement between the different 
formulations of the energy, and that the dispersive term provides
the Jacobian factor necessary to derive the Lifshitz free energy from the
expectation value of the electromagnetic energy.

The following point, related to the possibilities of experimental observations,
ought to be noticed: As we have seen, a characteristic property of dispersion
is that the factor $d(\varepsilon \omega)/d\omega$ occurs in the energy and
not in the pressure or the stress. This has a bearing on the famous
Abraham-Minkowski energy-momentum problem. As is known, an important experiment
in this area is the Jones-Richards radiation pressure experiment
\cite{jones54}, showing how the effective pressure against a mirror immersed in
a liquid varies with respect to the refractive index (cf. also the follow-up
experiment of Jones and Leslie \cite{jones78}). The book by Jones
\cite{jones88} contains a nice exposition of these very accurate experiments.
The electrostrictive forces do not contribute to the radiation pressure.

Does this imply that electrostrictive forces in a liquid are generally
non-observable? Not quite so, although a difficulty is that at thermal
equilibrium the electrostrictive forces give rise to elastic pressures in the
liquid, acting in the opposite direction. There are ways to overcome this
difficulty, however. One option is to proceed as in the Goetz-Zahn
non-equilibrium experiment \cite{goetz58,zahn62}; cf. also the detailed
discussion on this experiment in Ref.~\cite{brevik79}, p. 149.
One applies an electric field with high frequency $\omega$ between two
condenser plates in a liquid, and measures the attractive force between the
plates for instance by means of a piezoelectric transducer. The point is that
$\omega$ must be so high that the elastic pressure does not have time to built
itself up. The critical parameter here is thus the velocity of sound in the
liquid.

\acknowledgments
We thank the US Department of Energy, and the US National Science Foundation,
for partial support of this research.  We thank Elom Abalo for collaborative
assistance.

\appendix
\section{Constancy of energy for conducting plates}
Consider the case of parallel perfect conducting plates separated by a 
nondispersive medium with dielectric constant $\epsilon$.  The TE Green's
function is [obtained by taking $\kappa_{1,2}\to \infty$ 
in Eq.~(\ref{intgreen})]
\be
g^H=\frac1{2\kappa}\left\{e^{-\kappa|z-z'|}+\frac{2\cosh\kappa(z-z')
-e^{\kappa(z+z')}-e^{-\kappa(z+z'-2a)}}{e^{2\kappa a}-1}\right\},\ee
where $\kappa^2=k^2+\zeta^2\epsilon$.  The TE energy density is given
by
\be
u^{\rm TE}=\frac12\epsilon\langle E_y^2\rangle+\frac12\langle H_x^2+H_z^2
\rangle,
\ee
where
\begin{subequations}
\bea
\langle E_y^2\rangle&=&-\int_{-\infty}^\infty\frac{d\zeta}{2\pi}
\int\frac{(d\mathbf{k_\perp})}{(2\pi)^2}
\zeta^2 g^H(z,z),\\
\langle H_x^2+H_z^2\rangle&=& \int_{-\infty}^\infty\frac{d\zeta}{2\pi}
\int\frac{(d\mathbf{k_\perp})}{(2\pi)^2}
(\partial_z\partial_{z'}+k^2) g^H(z,z')
\big|_{z'=z}.
\eea
\end{subequations}
Then we easily see that
\be
u^{\rm TE}=\int_{-\infty}^\infty \frac{d\zeta}{2\pi}\int_0^\infty \frac{dk\,k}
{2\pi}\frac1\kappa\frac1{e^{2\kappa a}-1}\left[-\epsilon\zeta^2
-\frac{k^2}2\left(e^{2\kappa z}+e^{-2\kappa(z-a)}\right)\right].\label{ute}
\ee

The TM Green's function for perfectly conducting plates has a similar form,
\be
g^E=\frac1{2\kappa'}\left\{e^{-\kappa|z-z'|}+\frac{2\cosh\kappa(z-z')
+e^{\kappa(z+z')}+e^{-\kappa(z+z'-2a)}}{e^{2\kappa a}-1}\right\},\ee
and the corresponding energy density is
\bea
u^{\rm TM}&=&\frac12\epsilon\langle E_x^2+E_z^2\rangle +\frac12\langle H_y^2
\rangle\nonumber\\
&=&\int_{-\infty}^\infty\frac{d\zeta}{2\pi}
\int\frac{(d\mathbf{k_\perp})}{(2\pi)^2}
\left[\frac1{2\epsilon}(\partial_z
\partial_{z'}+k^2)g^E(z,z')\big|_{z'=z}\right.\nonumber\\
&&\qquad\qquad\left.\mbox{}-\frac1{2\zeta^2\epsilon^2}
(\partial_z^2-k^2)(\partial_{z'}^2-k^2)g^E(z,z')\big|_{z'=z}\right]\nonumber\\
&=&\int_{-\infty}^\infty \frac{d\zeta}{2\pi}\int_0^\infty \frac{dk\,k}{2\pi}
\frac1\kappa\frac1{e^{2\kappa a}-1}\left[-\zeta^2\epsilon+\frac{k^2}2\left(
e^{2\kappa z}+e^{-2\kappa (z-a)}\right)\right].\label{utm}
\eea

The $z$-dependent terms exactly cancel between Eqs.~(\ref{ute}) and 
(\ref{utm}) and the remaining terms are equal, and sum to the usual
Casimir energy density,
\be
u=-\frac1{3\pi^2\sqrt{\epsilon}}\int_0^\infty d\kappa\,\kappa^3
\frac1{e^{2\kappa a}-1}=-\frac{\pi^2}{720\sqrt{\epsilon}a^4}.
\ee

This cancellation, resulting in the constancy of the energy density, is
rather special, however.  It does not occur if dispersion is present,
$d\epsilon/d\zeta\ne0$, in which case the local energy density has the
nonconstant form:
\be
u=\int_{-\infty}^\infty\frac{d\zeta}{2\pi}
\int\frac{(d\mathbf{k_\perp})}{(2\pi)^2}\frac1\kappa
\frac1{e^{2\kappa a}-1}\left\{-\zeta^2\epsilon\left(2+\frac\zeta\epsilon\frac{
d\epsilon}{d\zeta}\right)+\frac{k^2}2
\frac\zeta\epsilon\frac{d\epsilon}{d\zeta}\left(e^{2\kappa z}
+e^{-2\kappa(z-a)}\right)\right\}.
\ee
Nor can the cancellation occur for dielectric media
constituting regions 1 and 2, since the TE and
TM reflection coefficients are then different.  Nevertheless, we note that the $z$
integral of the spatially varying part of the energy is a constant, independent
of $a$, and so does not contribute to the force on the plates.  This is
just as occurs for a nonconformally coupled massless scalar field confined
between Dirichlet plates.


\begin{thebibliography}{99}
\bibitem{Bordagbook} M. Bordag, G. L. Klimchitskaya, U. Mohideen, and V. M.
Mostepanenko, {\it Advances in the Casimir Effect} (Oxford Science 
Publications, Oxford, 2009).

\bibitem{Jaffe}
  R.~L.~Jaffe,
  Phys.\ Rev.\  D {\bf 72}, 021301 (2005)
  [arXiv:hep-th/0503158].

\bibitem{brevik08}
  I.~Brevik and K.~A.~Milton,
  Phys.\ Rev.\  E {\bf 78}, 011124 (2008)
  [arXiv:0802.2542 [quant-ph]].

\bibitem{landaulifshitz}
L. D. Landau and E. M. Lifshitz, {\it Electrodynamics of Continuous
Media}, 2nd ed. (Butterworth-Heinemann, Oxford, 1984).

\bibitem{schwingerbook} J. Schwinger, L. L. DeRaad, Jr., K. A. Milton, and
W.-y. Tsai, {\it Classical Electrodynamics} (Perseus, New York, 1998).

\bibitem{miltonbook} K. A. Milton, {\it The Casimir Effect: Physical
Manifestations of Zero-Point Energy} (World Scientific, Singapore, 2001).

\bibitem{sdm}
  J.~Schwinger, L.~L.~DeRaad, Jr., and K.~A.~Milton,
  Ann.\ Phys.\ (N.Y.)  {\bf 115}, 1 (1978).

\bibitem{renne}
M. J. Renne, Physica {\bf 56}, 125 (1971).

\bibitem{reynaud}
A. Lambrecht, P. A. Maia Neto, and S. Reynaud,
 New J. Phys. {\bf8}, 243 (2006).

\bibitem{miltonrev}
  K.~A.~Milton,
  J.\ Phys.\ A  {\bf 37}, R209 (2004)
  [arXiv:hep-th/0406024].



\bibitem{brevikfest}
K. A. Milton,
P. Parashar and J. Wagner, in {\it The Casimir Effect and Cosmology}, 
ed. S. D. Odintsov, E. Elizalde, and O. B. Gorbunova, 
in honor of Iver Brevik (Tomsk State Pedagogical University) pp. 107-116 
(2009) [arXiv:0811.0128].

\bibitem{brown}
  L.~S.~Brown and G.~J.~Maclay,
  Phys.\ Rev.\  {\bf 184}, 1272 (1969).


\bibitem{jones54}
R. V. Jones and J. C. Richards, Proc. Roy. Soc. London Ser. A {\bf 221}, 480 (1954).
\bibitem{jones78}
R. V. Jones and B. Leslie, Proc. Roy. Soc. London Ser. A {\bf 360}, 347 (1978).
\bibitem{jones88}
R. V. Jones, {\it Instruments and Experiences} (John Wiley \& Sons, New York, 1988).
\bibitem{goetz58}
H. Goetz and W. Zahn, Zeitschr. Physik {\bf 151}, 202 (1958).
\bibitem{zahn62}
W. Zahn, Zeitschr. Physik {\bf 166}, 275 (1962).
\bibitem{brevik79}
I. Brevik, Physics Reports {\bf 52}, 133 (1979).

\end{thebibliography}
\end{document}